\documentclass{aa}  
\newcommand{\HII}{H\,{\scriptsize II}}

\newcommand{\HI}{H\,{\scriptsize I}}
\newcommand{\CII}{[C\,{\scriptsize II}]}

\newcommand{\hi}{H\,{\scriptsize I}}

\newcommand{\CI}{[C\,{\scriptsize I}]}

\newcommand{\av}{A$_{\rm V}$}

\usepackage{graphicx}  
\usepackage{txfonts}
\usepackage{caption}
\usepackage[colorlinks=true,     linkcolor=blue, citecolor=blue, filecolor=blue, urlcolor=blue]{hyperref}

\begin{document} 

\title{The HI-to-H$_2$ transition in the Draco cloud}

\author{Nicola Schneider \inst{1}
\and Volker Ossenkopf-Okada  \inst{1} 
\and Markus R\"ollig  \inst{2,1}
\and Daniel Seifried \inst{1}  
\and Ralf S. Klessen \inst{3}
\and Alexei G. Kritsuk \inst{4}
\and Eduard Keilmann \inst{1} 
\and Simon Dannhauer \inst{1}
\and Lars Bonne \inst{5}
\and Simon C.O. Glover \inst{3} 
}

\institute{I. Physikalisches Institut, Universit\"at zu K\"oln, Z\"ulpicher Str. 77, 50937 K\"oln, Germany  
\email{nschneid@ph1.uni-koeln.de}
\and Physikalischer Verein, Gesellschaft für Bildung und Wissenschaft, Robert-Mayer-Str. 2, 60325 Frankfurt, Germany 
\and Universit\"at Heidelberg, Zentrum f\"ur Astronomie, 69120 Heidelberg, Germany 
\and Physics Department, University of California, San Diego, La Jolla, CA 92093-0319, USA 
\and SOFIA Science Center, NASA Ames Research Center, Moffett Field, CA 94 045, USA 
}

\date{draft of \today}

\titlerunning{HI to H$_2$ transition in Draco}  
\authorrunning{N. Schneider}  
  
\abstract{ In recent decades, significant attention has been dedicated
  to analytical and observational studies of the atomic hydrogen (\HI)
  to molecular hydrogen (H$_2$) transition in the interstellar
  medium. We focussed on the Draco diffuse cloud to gain deeper
  insights into the physical properties of the transition from \HI\ to
  H$_2$.  We employed the total hydrogen column density probability
  distribution function (N-PDF) derived from \textit{Herschel} dust
  observations and the N$_{\rm HI}$-PDF obtained from \HI\ data
  collected by the Effelsberg \HI\ survey.  The N-PDF of the Draco
  cloud exhibits a double-log-normal distribution, whereas the N$_{\rm
    HI}$-PDF follows a single log-normal distribution. The
  \HI-to-H$_2$ transition is identified as the point where the two
  log-normal components of the dust N-PDF contribute equally; it
  occurs at A$_{\rm V} \sim 0.33$ (N $\sim 6.2 \times 10^{20}$
  cm$^{-2}$).  The low-column-density segment of the dust N-PDF
  corresponds to the cold neutral medium, which is characterized by a
  temperature of around 100 K. The higher-column-density part is
  predominantly associated with H$_2$. The shape of the Draco N-PDF is
  qualitatively reproduced by numerical simulations.  In the absence
  of substantial stellar feedback, such as radiation or stellar winds,
  turbulence exerts a significant influence on the thermal stability
  of the gas and can regulate the condensation of gas into denser
  regions and its subsequent evaporation. Recent observations of the
  ionized carbon line at 158 $\mu$m in Draco support this
  scenario. Using the KOSMA-tau photodissociation model, we estimate a
  gas density of n $\sim 50$ cm$^{-3}$ and a temperature of $\sim 100$
  K at the location of the \HI-to-H$_2$ transition. Both the molecular
  and atomic gas components are characterized by supersonic turbulence
  and strong mixing, suggesting that simplified steady-state chemical
  models are not applicable under these conditions.  }

   \keywords{ISM:dust, extinction - ISM:clouds - ISM:structure}

   \maketitle
 
\section{Introduction} \label{sec:intro}

The formation of molecular hydrogen (H$_2$) in the interstellar medium
(ISM) has historically been described using steady-state and chemical
equilibrium models
\citep{Tielens1985,Ewine1988,Sternberg1989,Krumholz2008}. In such
models, the dissociation of H$_2$ by photons in the Lyman-Werner bands
from the local radiation field competes with H$_2$ formation on the
surfaces of dust grains. Although H$_2$ self-shielding and dust shielding against UV photodissociation are highly effective, the
formation rate of molecular H$_2$ remains relatively low,
approximately $3 \times 10^{-17}$~cm$^3$ s$^{-1}$ \citep{Jura1974}.
The H$_2$ formation rate and the efficiency of dust shielding are
further influenced by the abundance of dust grains, which is commonly
scaled in models based on metallicity. For a more detailed discussion
of the atomic-to-molecular hydrogen (\HI-to-H$_2$) transition and
associated references, see \citet{Bellomi2020} and \citet{Park2023}.
In dynamic scenarios, however, H$_2$ formation is also influenced by
turbulent mixing motions in the ISM
\citep{Glover2007,Bialy2017}. These motions induce large- and
small-scale density fluctuations that can drastically reduce H$_2$
formation timescales from tens of millions of years to just a few
million years \citep{Glover2007,Valdivia2016}. This reduction in
timescales has profound implications for the ISM. For instance,
\citet{MacLow2012} argue that star-forming clouds are often disrupted
by stellar feedback before they can reach chemical equilibrium.

Turbulence in the ISM can be generated by a variety of
mechanisms. These include colliding \HI\ flows and other dynamic
processes that lead to H$_2$ formation in shock-compressed layers
\citep{Walder1998,Koyama2000,Heitsch2006,Vaz2006,Dobbs2008,Hennebelle2008,
  Banerjee2009,Clark2012,Colman2025}. Additionally, large-scale
gravitational instabilities in the galactic disc and stellar feedback,
predominantly from supernova explosions, contribute significantly to
turbulence generation \citep{MacLow2004,Klessen2016}.

\begin{figure*}
\centering
\includegraphics[width=9cm,angle=0]{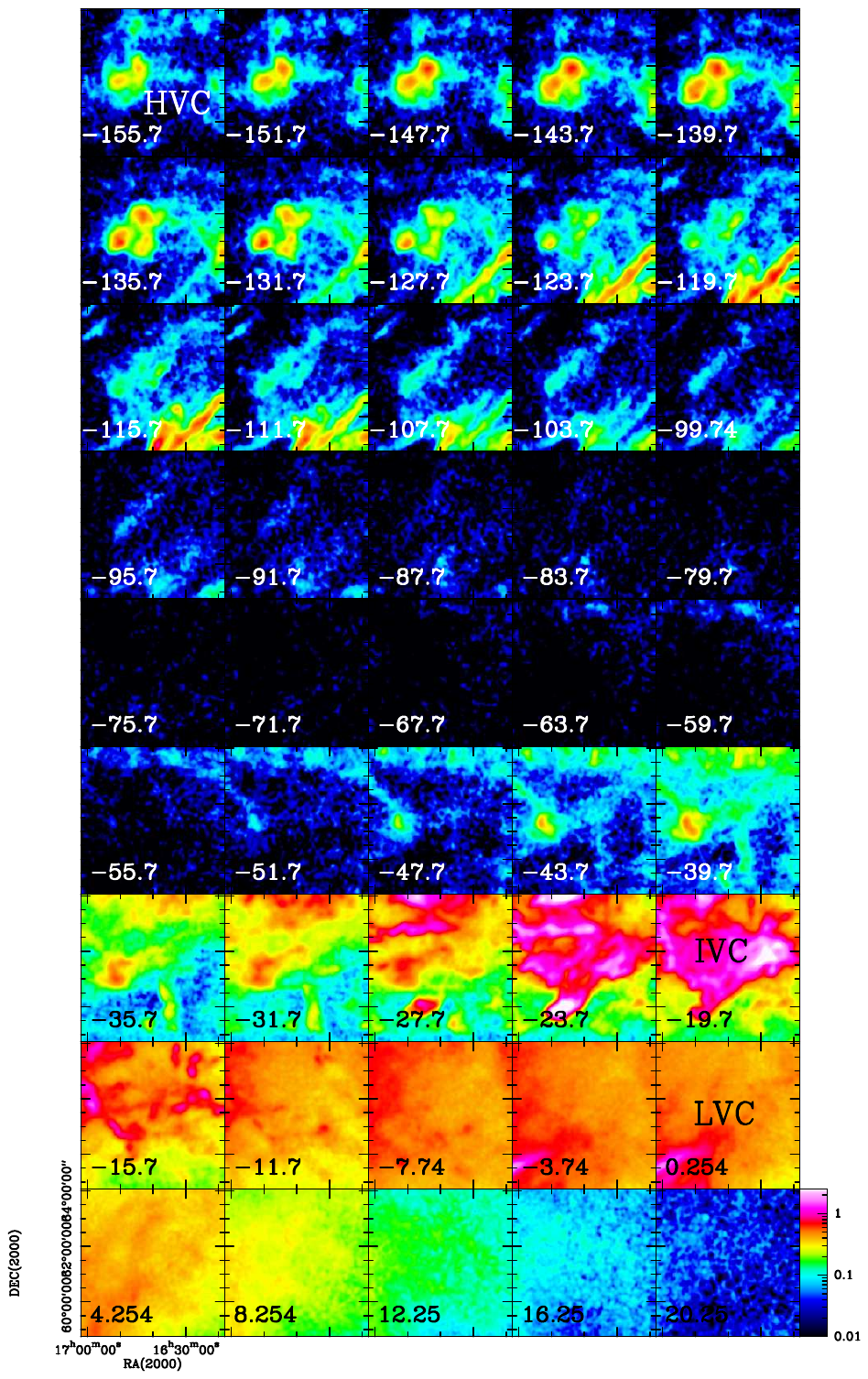}
\includegraphics[width=9cm,angle=0]{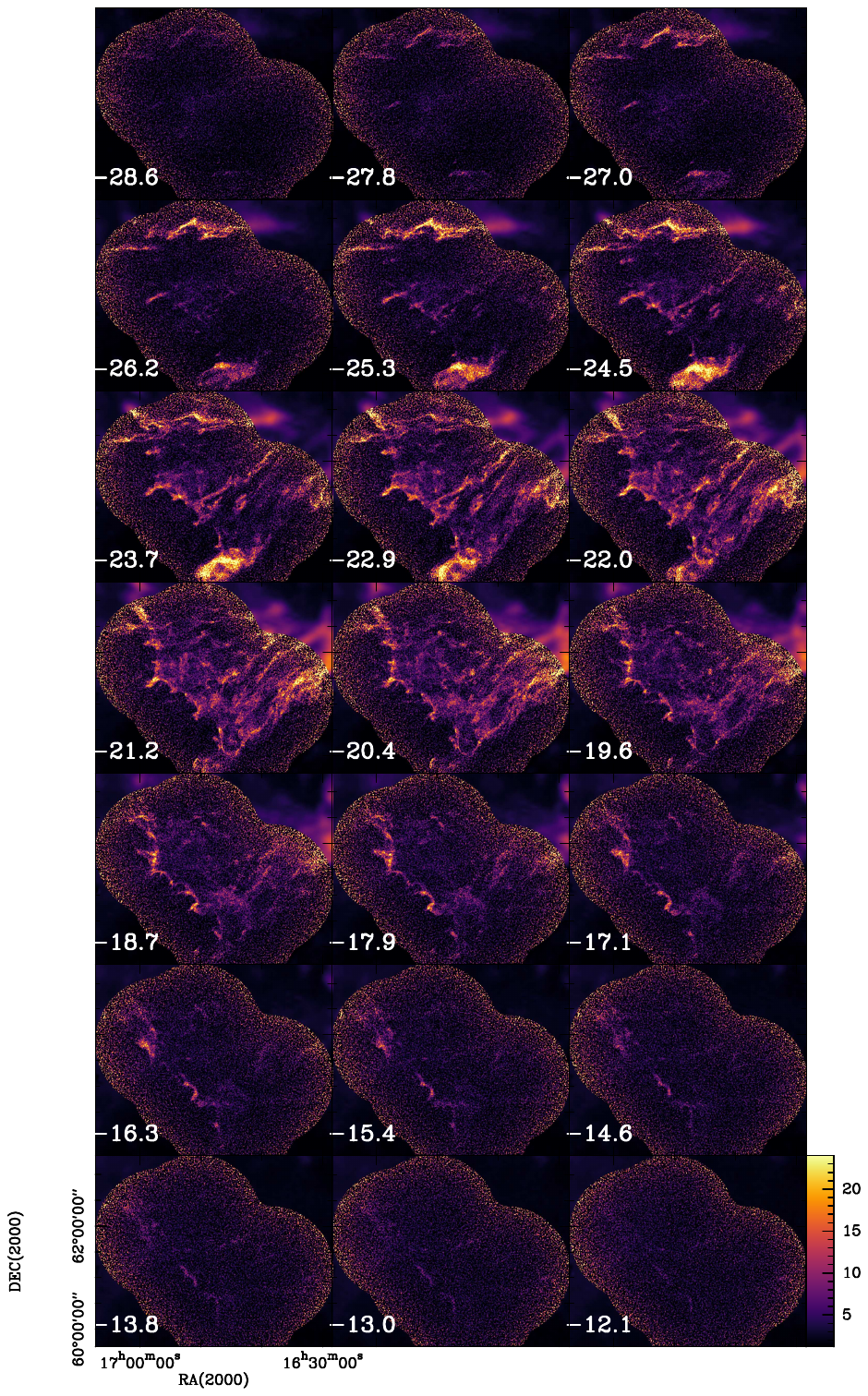}
\caption{Draco cloud in \HI. Left: Channel maps of
  \HI\ brightness temperature in K from EBHIS at 10$'$ angular
  resolution in steps of 4 km s$^{-1}$. The central velocity is given
  in the lower-left corner, and the individual velocity components of
  Draco (high, intermediate, and local) are indicated.  Right: Zoom into the velocity range of the IVC between -12 and
  -28 km s$^{-1}$ using the DRAO with the Synthesis Telescope
  \HI\ data at 1$'$ angular resolution \citep{Blagrave2017} in steps
  of 0.82 km s$^{-1}$.} \label{draco-HI}
\end{figure*}

In plane-parallel photodissociation region (PDR) models
\citep{Ewine1986,Hollenbach1991,Wolfire2010,Sternberg2014}, the
\HI-to-H$_2$ transition is parameterized by the density of the gas and
the intensity of the incident radiation field. For low densities
(typically $n < 10^3$ cm$^{-3}$) and a weak to moderate
far-ultraviolet (FUV) radiation field,\footnote{The FUV field is
expressed in units of Habing G$_{\rm o}$ \citep{Habing1968} or Draine
$\chi$ \citep{Draine1978}, where $\chi$ = 1.71 G$_{\rm o}$.} with $\chi$
in the range 1–10$^3$, the \HI-to-H$_2$ transition occurs at visual
extinctions (\av) below unity.\footnote{We used the conversion $N_{\rm
  H}$/\av = 1.87 $\times 10^{21}$ cm$^{-2}$ mag$^{-1}$
\citep{Bohlin1978} to relate the total hydrogen column density to the
visual extinction.}  Due to molecular formation and the cumulative
dust shielding, this transition is accompanied by a significant drop
in temperature, typically from values exceeding 100 K to approximately
20 K. \citet{Imara2016} observed a column density of
$N_{\rm H} \sim 8$–$20 \times 10^{20}$ cm$^{-2}$ (\av$\sim 0.4$–$1.1$)
for the \HI-to-H$_2$ transition in optically thick molecular clouds
surrounded by \HI\ envelopes. Studies of infrared cirrus, as well as
intermediate-velocity clouds (IVCs) and high-velocity clouds
\citep[HVCs:][]{Reach1994,Lagache1998,Gillmon2006,Gillmon2006b,Roehser2014},
suggest values of $N_{\rm H} \sim 2$–$5 \times 10^{20}$ cm$^{-2}$
(\av$\sim 0.1$–$0.3$) for the \HI-to-H$_2$ transition.  Similar values
were found by \citet{Schneider2022} using column density probability distribution
functions (N-PDFs) of dust for diffuse, quiescent, and
low-density clouds. Moreover, the \HI\ to far-infrared (FIR) continuum
relation turns non-linear at values above $N_{\rm H} = 5 \times
10^{20}$ cm$^{-2}$ \citep{Desert1988,Lenz2015,PlanckXXIV2011}, which
is interpreted as the transition from atomic to molecular hydrogen.

While the chemical state of the ISM is governed by the balance between
photodissociation and H$_2$ formation on dust grains, its thermal
state is regulated by heating processes, such as the photoelectric
effect, and cooling, primarily through collisionally excited emission
in FIR molecular and atomic fine-structure lines. The relatively
shallow dependence of the cooling rate on temperature for T $<$ 10$^4$
K, dominated by cooling via the ionized carbon \CII\ 158 $\mu$m line,
leads to thermal instabilities that produce a multi-phase ISM
\citep{Field1969,Wolfire1995}.  It consists of volume-filling warm
neutral medium (WNM), characterized by temperatures of T$\sim$ 8000 K
and densities $n \sim 1$ cm$^{-3}$, in pressure equilibrium with the
cold neutral medium (CNM), which has temperatures of T $\sim$ 30–100 K
and densities $n \sim$ 10–100 cm$^{-3}$
\citep{wolfire2003,Bialy2019}. Additionally, molecular H$_2$ gas
typically exhibits temperatures below 30 K and densities exceeding
several hundred cm$^{-3}$. However, thermally unstable gas, often
referred to as the lukewarm neutral medium or the unstable neutral
medium \citep[UNM;][]{Murray2018}, can also exist with temperatures
in between those of the CNM and the WNM \citep{Marchal2019}. Depending on
its density, such gas can cool to become denser and fully
molecular, or it can heat up and join the WNM
\citep{Heiles2003}. Under turbulent conditions, significant amounts of
thermally unstable gas can be present as turbulent motions expand the
interface between stable thermal phases. Although this unstable gas is
challenging to observe, it plays a critical role in probing the
\HI-to-H$_2$ transition as a function of the hydrogen column density and
the incident UV field, offering insights into the formation of
molecular clouds and stars from diffuse \HI\ gas. A possible tracer of
this gas phase is the \CII\ line, as proposed by \citet{SChneider2023},
who observed interacting flows -- partly atomic, partly molecular -- in
the Cygnus X region. This gas, which does not emit prominently in
the lines of carbon monoxide and is thus called `CO-dark'
\citep{Wolfire2010}, has a density of $n \sim 100$ cm$^{-3}$ at T
$\sim$ 100 K in Cygnus.

To investigate the \HI-to-H$_2$ transition, we selected the Draco
IVC, a region that is likely on the
verge of forming a molecular cloud and is exposed to a very low incident
UV field. The dust column density map of this high-latitude cloud was obtained from FIR dust continuum maps based on observations taken with
the \textit{Herschel} satellite and was presented in
\citet{Schneider2022, Schneider2024}. This map, which spans a large
dynamic range, includes both atomic and molecular phases, with the
relative contributions of these phases depending on the cloud's
evolutionary state. Other column density maps of Draco at various
angular resolutions obtained from \textit{Herschel} were created by
\citet{Deschenes2017} and \citet{Bieging2024}.  In
\citet{Schneider2022} we demonstrate that the atomic and molecular
gas in Draco can be separated using N-PDFs. For the atomic N-PDF, we
utilized \HI\ data from the Effelsberg \HI\ survey \citep{Winkel2016},
with an angular resolution of 10$'$. In this study we investigated the physical conditions (e.g. the density and temperature) and
dynamics (e.g. the turbulent Mach number) of the atomic and molecular
gas. We show that turbulent mixing significantly influences the
chemical evolution of clouds and that steady-state equilibrium models
fail to explain the observed gas properties.

We start with a short description of the source
(Sect.~\ref{sec:source}) and a summary of the interpretation of the
N-PDFs. We then derive the physical properties of the gas using a PDR
model, describe its turbulent properties, and compare the observed
N-PDF with those obtained from simulations (Sect.~\ref{sec:analysis}).
Section~\ref{sec:discuss} concludes the paper.

\begin{figure}
\hspace{-1cm} \includegraphics[width=13cm,angle=0]{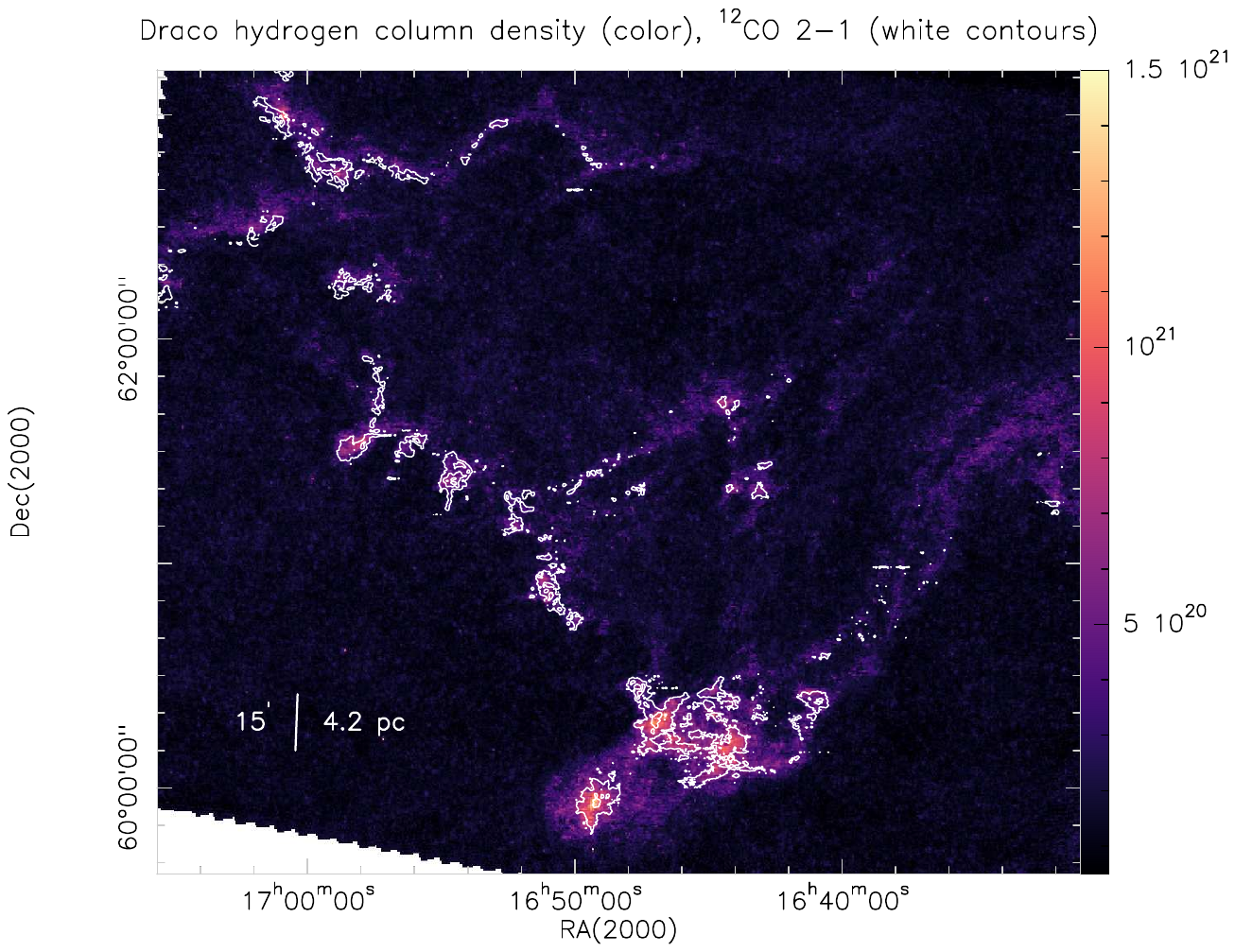}
\caption{Draco total hydrogen column density map (N(H)) and CO 2$\to$1
  contours.  The \textit{Herschel} N(H) map
  \citep{Schneider2022,Schneider2024} has an angular
  resolution of 36$''$. The $^{12}$CO 2$\to$1 data have a resolution
  of 38$''$ and stem from \citet{Bieging2024}. The contours are 2, 7,
  and 12 K km s$^{-1}$.  } \label{draco-map}
\end{figure}

\section{The Draco cloud} \label{sec:source} 

The Draco cloud (IVC G091.0+38.0, MBM41) is situated above the
Galactic plane \citep{Gladders1998} at a Galactic latitude of
$b=38^{\circ}$, corresponding to a scale height of 298 pc, assuming a
distance of 481~pc.\footnote{The distance to the IVC was initially
estimated to be between 463 and 618~pc (with an uncertainty of
$\sim$200~pc) by \citet{Gladders1998}, using sodium doublet absorption
measurements. Here, we adopt the more recent distance of $481 \pm
50$~pc, determined by \citet{Zucker2020} based on Gaia DR2 parallax
measurements.} Draco is classified as an IVC with a radial velocity of approximately $-20$~km s$^{-1}$. It is
part of the IR cirrus and is associated with diffuse \HI\ emission
\citep{Heiles1974}.  The Draco cloud is thought to originate from a
Galactic fountain process, in which material from the Galactic disc is
lifted into the halo and subsequently returns to the disc at high
velocities \citep{Lenz2015} and references therein. However, the
possibility of infalling extragalactic gas contributing to its origin
cannot be entirely ruled out. The Draco IVC is located within the
inner Galactic halo and thus outside of the bulk of the WNM in the
Galactic disc.

In the HVC part of the region there is gas with significantly higher
negative velocities ($v \sim -100$~km s$^{-1}$), and in the
low-velocity cloud (LVC) part there are local velocities near 0 km
s$^{-1}$. Figure~\ref{draco-HI} presents \HI\ emission channel
maps spanning the entire velocity range, from that of the LVC to that
of the HVC, alongside a zoomed-in view of the IVC velocity
range. Analysis of the low-angular-resolution ($10'$) Effelsberg-Bonn
\HI\ Survey (EBHIS) data reveals the absence of a velocity bridge
connecting the HVC and IVC components. Additionally, there is a
notable gap in emission around $-70$~km s$^{-1}$, meaning there is no
clear evidence that Draco originates from a collision between the HVC
and IVC, as was previously hypothesized by \citet{Herbstmeier1993}.
Higher-angular-resolution ($1'$) data from the Dominion Radio
Astrophysical Observatory (DRAO) highlight the small-scale structure
of the Draco cloud. At velocities beginning near $-13$~km s$^{-1}$, a
broad front emerges, characterized by individual knots of emission. At
more negative velocities ($-19$~km s$^{-1}$), filamentary structures
become apparent, oriented towards the Galactic plane. This morphology
was interpreted by \citet{Kalberla1984}, \citet{Rohlfs1989}, and
\citet{Deschenes2017} as indicative of gas compression resulting from
a halo cloud interacting with a more diffuse medium as it moves
towards the Galactic plane.  We note that this paper does not focus on
the general physics of IVCs. For further discussion on IVCs, we refer
the reader to \citet{Putman2012}, \citet{Roehser2014},
\citet{Roehser2016a}, \citet{Roehser2016b}, \citet{Kerp2016}, and
references therein.

The hydrogen column density map obtained with \textit{Herschel} is
presented in Fig.~\ref{draco-map}. See Appendix~\ref{app-cib} for
further information on this dust column density map. In the regions of
highest column density, CO and other molecules with critical densities
exceeding $10^3$ cm$^{-3}$ have been detected
\citep{Mebold1985,Herbstmeier1993, Schneider2024}.
\citet{Bieging2024} present a $^{12}$CO 2$\to$1 map of the whole Draco
region at 38$''$ angular resolution. Comparing our total hydrogen
column density map with the CO data (Fig.~\ref{draco-map}), it becomes
obvious that CO is found in the densest regions with a column density
around 0.8-1 10$^{21}$ cm$^{-2}$. These dense clumps consist of
molecular hydrogen (H$_2$) and are CO-bright. The remaining emission
in the \textit{Herschel} map is then either CO-dark molecular gas or
atomic gas. Note that the CO-bright H$_2$ clumps are also visible in
\HI\ emission, as shown in the right panel of
Fig.~\ref{draco-HI}. \citet{Bieging2024} derived an H$_2$ map by
subtracting the DRAO \HI\ map from a total \textit{Herschel} hydrogen
column density map that was obtained using only the 250 $\mu$m
emission (and not by a spectral energy distribution fit as done in
\citealt{Schneider2022}). With this method, they obtained that CO
formation starts at a column density of around 1.3 10$^{21}$ cm$^{-2}$.
\citet{Bieging2024} estimated that a high fraction of gas in Draco is
molecular, in contrast to \citet{Schneider2022}, who derived that
$\sim$89\% of the mass in the Draco cloud is composed of atomic
hydrogen.  \citet{Deschenes2017}, using \textit{Herschel} dust
observations, concluded that the molecular gas primarily consists of
small clumps with characteristic sizes of $\sim$0.1 pc, high densities
($n \sim 1000$ cm$^{-3}$), and low temperatures (T $\sim$ 20 K). The
highly clumpy structure of the gas was confirmed by the CO map of
\citet{Bieging2024}. However, the Draco cloud exhibits no evidence of
ongoing star formation, as no pre-stellar or protostellar cores have
been detected.

Recently, \citet{Schneider2024} reported the first detection of the
158 $\mu$m \CII\ line at five positions in the Draco IVC. Combining CO,
\HI, and \CII\ data led us to the conclusion that shocks heat the gas
that subsequently emit in the \CII\ cooling line. These shocks are
probably caused by the motion of the cloud towards the Galactic plane
that leads to collisions between \HI\ clouds.

\section{Analysis and discussion} \label{sec:analysis}

In the following analysis, we first determine the physical properties
of the gas in the Draco cloud using a PDR
model (Sect.~\ref{subsec:prop}). Subsequently, we examine the
turbulent properties of the multi-phase gas in Draco
(Sect.~\ref{subsec:turb-prop}) and then compare the observed Draco
N-PDF to those obtained from simulations
(Sect.~\ref{subsec:simu}). For this analysis, we utilized the dust and
\HI\ N-PDFs
presented in \citet{Schneider2022} and depicted in
Fig.~\ref{pdf}. This figure shows a double log-normal PDF for the
total hydrogen column density. The lower-column-density component,
characterized by a width of $\sigma_N = 0.32$, is attributed to atomic
gas, while the higher-column-density component, with a width $\sigma_N
= 0.34$, is associated with molecular gas. The \HI-to-H$_2$ transition
occurs where the contributions from the two log-normal components are
equal, corresponding to an extinction value of \av = 0.33 or a column
density of $N_{\rm H}$ = 6.2 $\times$ 10$^{20}$ cm$^{-2}$. Note,
however, that H$_2$ can also occur already at lower column densities,
but with a lower probability. \citet{Bieging2024} measured an \av\ of
0.7 for the onset of CO formation. This value is lower than typical CO
formation scale thresholds \citep{lee1996,visser2009}, but well
explained if a small-scale clumpy structure of the molecular material
is considered with dense clumps of size $\leq$0.1 pc. We come
back to this point in Sect.~\ref{subsec:prop}.

Column density PDFs are widely employed as diagnostic tools to
characterize the physical processes that shape the column density
structure of the ISM. It has been shown that a log-normal density
distribution emerges naturally for isothermal flows
\citep[e.g.][]{Passot1998,Vaz1994,Padoan1997,Klessen2000,Li2003,Kritsuk2007,Federrath2008,Ball2011,Burkhart2012}
and for thermally bistable flows \citep{Audit2010,Gazol2013}, as a
result of the central limit theorem. This assumes that density
perturbations caused by successive shock passages accumulate
independently and randomly.

Previous \textit{Herschel} studies of evolved star-forming regions have
consistently shown a log-normal component in the N-PDF at low column
densities. These are often accompanied by one
\citep[e.g.][]{Schneider2013,Schneider2015a,Schneider2015b,Stutz2015}
or two \citep{Schneider2015c,Schneider2022} power-law tails at higher
column densities, which are typically attributed to the effects of
self-gravity
\citep[e.g.][]{Kritsuk2011,Girichidis2014,Schneider2015a,Schneider2015b}. For
example, N-PDFs of the Lupus and Coalsack clouds derived from
extinction maps \citep{Kainulainen2009} appear log-normal in
shape. However, \textit{Herschel} dust-based N-PDFs, which provide a
higher dynamic range, reveal a log-normal component at low column
densities followed by a power-law tail at higher column densities
\citep{Benedettini2015}. Double log-normal N-PDFs with a single
power-law tail at high column densities were observed by
\citet{Tremblin2014} for molecular clouds associated with
\HII\ regions. In these star-forming clouds, the second log-normal
component is thought to be produced by feedback processes
\citep{Schneider2022}. For Draco, however, no power-law tail is
observed, indicating that self-gravity does not yet play a significant
role. As said above, this is observationally supported by the
non-detection of signposts of ongoing star formation. This suggests
that Draco represents an excellent example of a turbulence-dominated,
multi-phase region, where the total hydrogen column density is
primarily composed of atomic hydrogen.

\begin{figure}
\vspace{-1.5cm}
\hspace{-3cm}\includegraphics[width=15cm, angle=0]{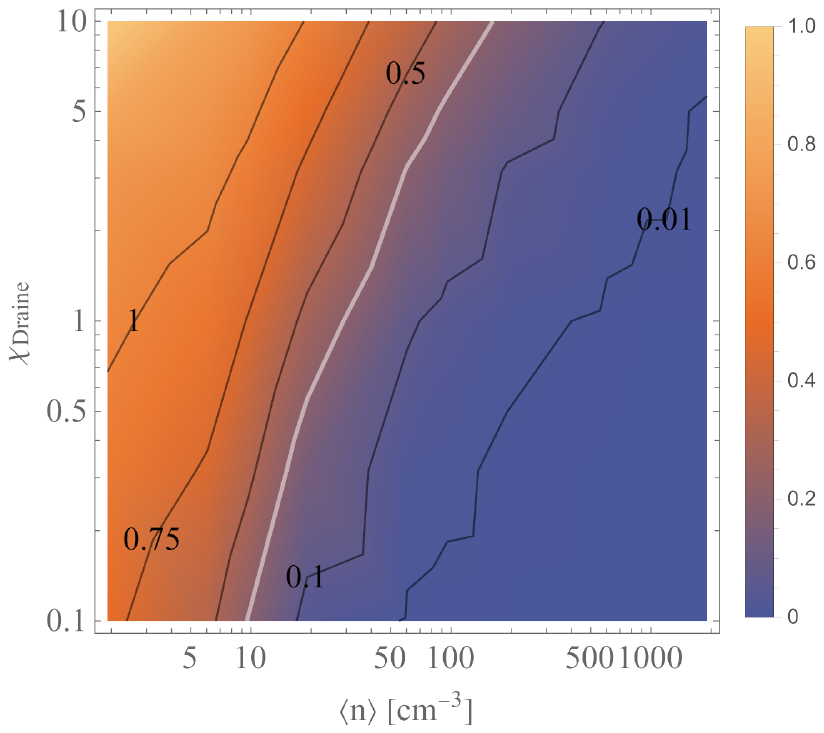}
\vspace{-2cm}
\caption{\HI-to-H$_2$ transitional \av\ as a function of UV field and
  the clump-averaged density $\langle n \rangle$, determined from the
  KOSMA-tau PDR model. The colour wedge gives the value of \av, and the
  white line indicates our observed transitional \av\ of 0.33.}
\label{av}
\end{figure}

\subsection{Physical properties of the gas} \label{subsec:prop}

To determine the conditions for the \HI-to-H$_2$ transition, we did not
try to reproduce the emission and clumpy nature of the Draco cloud
but instead performed the simple geometrical study measuring the critical depth
for H$_2$ formation in a plane-parallel configuration. The balance
between photodissociation of H$_2$ and its formation on dust grain
surfaces is controlled by the local gas density and the FUV photon
flux.

The FUV radiation field is computed by converting the 160 $\mu$m flux
observed in Draco into a FUV field assuming that the dust is only
heated by FUV radiation \citep{Schneider2024}. This gives a value of
$\chi\sim$1.8$-$2.5 at the dust peaks.\footnote{We note that the PACS
160 $\mu$m data were successfully obtained, but the 100 $\mu$m data
were lost due to a Signal Processing Unit (SPU) anomaly during this
observation day OD807 (see \textit{Herschel} archive quality control
report summary).}  This is higher than expected as there is no known
OB star close to the Draco cloud.  In \citet{Schneider2024}, we show
that cosmic rays are also not a sufficient heating source, which
leaves as the most likely reason shock excitation.  We assumed an
uncertainty of at least a factor of 2 and calculate with an average
field of $\chi\sim$2.

We used the KOSMA-tau PDR model
\citep{Roellig2006,Roellig2007,Roellig2022} to estimate the average
density, $\langle n \rangle$, as a function of the incident FUV field
and the \HI-to-H$_2$ transitional \av. The model is designed to
represent clouds by a superposition of spherical clumps with a given
mass and density or to model a single clump. Here, $n=n({\rm H}) + 2
n({\rm H_2})$ is the total proton density in cm$^{-3}$ and $\langle
\rangle$ indicates the average over the clumps.\footnote{The assumed
density profile in KOSMA-tau follows a power law with power-law index
$-3/2$ and a constant density if $r\le 0.2 R_\mathrm{tot}$. Under
these conditions $\langle n \rangle \approx 1.91 n_0$ where $n_0$ is
the total gas density at the surface of the clump.} In order to
approximate a plane-parallel situation in KOSMA-tau, we used a model
mass of M = $10^{6}$\,M$_\odot$; this resulted in very large cloud
radii. However, the total mass has no significant influence on the
\HI-to-H$_2$ transition. We computed a grid of model PDRs, covering
the diffuse to moderate gas parameter space through gas densities $n_0
= 1-1000$ cm$^{-3}$ and FUV intensities of $\chi=0.1-10$.  We did not
consider turbulent mixing (which is discussed in the next
section); we assumed isotropic illumination and expressed our results in
terms of A$_{\rm V,eff}$ \citep{Roellig2007} to be able to convert them from an
isotropic illumination to an unidirectional geometry. In
Fig.~\ref{av}, the depth for the transition from \HI\ to H$_2$, i.e.
the effective \av\ at which $n({\rm H})=n({\rm H_2})$, is plotted. The
general behaviour is intuitive: For a given model density, the
transition depth increases with $\chi$. For a fixed FUV strength, the
transition occurs at lower \av\ when the density grows.  For the
observed transitional \HI-to-H$_2$ \av\ of 0.33 and an UV-field of
G$_{\rm 0}=3.4$, equivalent to $\chi=2$, we then derive an average
density of $\langle n \rangle \sim$50 cm$^{-3}$. This density is
consistent with the mean density of 76~cm$^{-3}$ for cold \hi{}
filament gas at high Galactic latitudes derived by
\citet{Kalberla2025}.  Thus, the scenario we obtain is that small,
dense (up to a few times 10$^3$ cm$^{-3}$), cold (T$\sim$20 K) molecular
cloud clumps are embedded in a lower-density ($<$50 cm$^{-3}$)
inter-clump medium. An alternative to the steady-state solution
computed by KOSMA-tau might be higher densities combined with short
timescales. When starting from atomic material, the transition that
would occur at \av$\sim$0.03 for $\langle n\rangle = 300$~cm$^{-3}$ in
a steady state may well occur at \av$\sim$0.3 on the cloud formation
timescale, which is too short for the system to reach chemical equilibrium.

\subsection{Turbulence properties of the gas in Draco} \label{subsec:turb-prop}

Dynamical effects in molecular cloud formation simulations are often
neglected and chemistry assumed to be at equilibrium in most of the
cases.  Exceptions are, for example, the studies of
\citet{Godard2014}, \citet{Valdivia2016}, \citet{Seifried2022}, and
\citet{Ebagezio2023}.  Because the dynamical timescale, $t_{dyn}$, is
inverse to the (density-independent) Mach number and the chemical timescale, $t_{chem}$, for H$_2$ formation on grains is inverse to the
density, we are closer to equilibrium when the gas is denser.  Because
of the relatively low densities in Draco we need to investigate the
turbulence properties to assess the importance of dynamical processes
in Draco, using the dust PDF \citep{Schneider2022} and CO observations
\citep{Mebold1985,Herbstmeier1993,Schneider2024,Bieging2024}.

Generally, for a log-normal density ($\rho$) PDF, the rms sonic Mach
number,\footnote{Strictly speaking, $\mathcal{M}$ is a parameter to
characterize turbulence. It is a measure of compressibility in a given
velocity field, and high values of $\mathcal{M}$ imply strong
compressibility and produce strong local density enhancements.}
$\mathcal{M}$ is linked to the standard deviation $\sigma_\rho$ of the
PDF via
\begin{equation} 
\sigma^2_\rho \, = \, \ln{(1+b^2 \mathcal{M}^2)}.  
\end{equation} 
The forcing parameter $b$ is related to the kinetic energy injection mechanism and varies from $b\approx$1/3 for solenoidal forcing to $b$=1 for compressive forcing \citep{Federrath2008}.  When magnetic fields are included, the density variance additionally depends on the ratio of thermal pressure $p_{th}$ to magnetic pressure $p_{mag}$
defined as
\begin{equation} 
\beta \, = \, \frac{p_{th}}{p_{mag}} \, = \, 2 \frac{c_s^2}{{\rm v}_A^2} 
,\end{equation}
with the sound speed $c_s$ and the Alfv\'enic velocity, ${\rm v}_A$. \citet{Molina2012} found that if the magnetic
field, $B,$ is proportional to $\rho^{1/2}$, the density variance is
\begin{equation} 
\sigma^2_{\rho,mag} \, = \, \ln{(1 + b^2 \mathcal{M}^2 \, \beta/(\beta+1))}.  
\end{equation}
For N-PDFs, \citet{Burkhart2012} derived the correlation
\begin{equation} 
\sigma^2_N \, = \, 0.11 \, \ln{(1+b^2 \mathcal{M}^2)}, 
\end{equation}
which does not include any magnetic field dependence. 
We assumed that the magnetic pressure term scales in a similar fashion as determined by \citet{Molina2012} (Eq.~3),  so
\begin{equation}
\sigma^2_{N,mag} \, = \, 0.11 \, \ln{(1 + b^2 \mathcal{M}^2 \, \beta/(\beta+1))}. \label{eq:b}
\end{equation}

\begin{figure}
\centering
\includegraphics[width=8cm, angle=0]{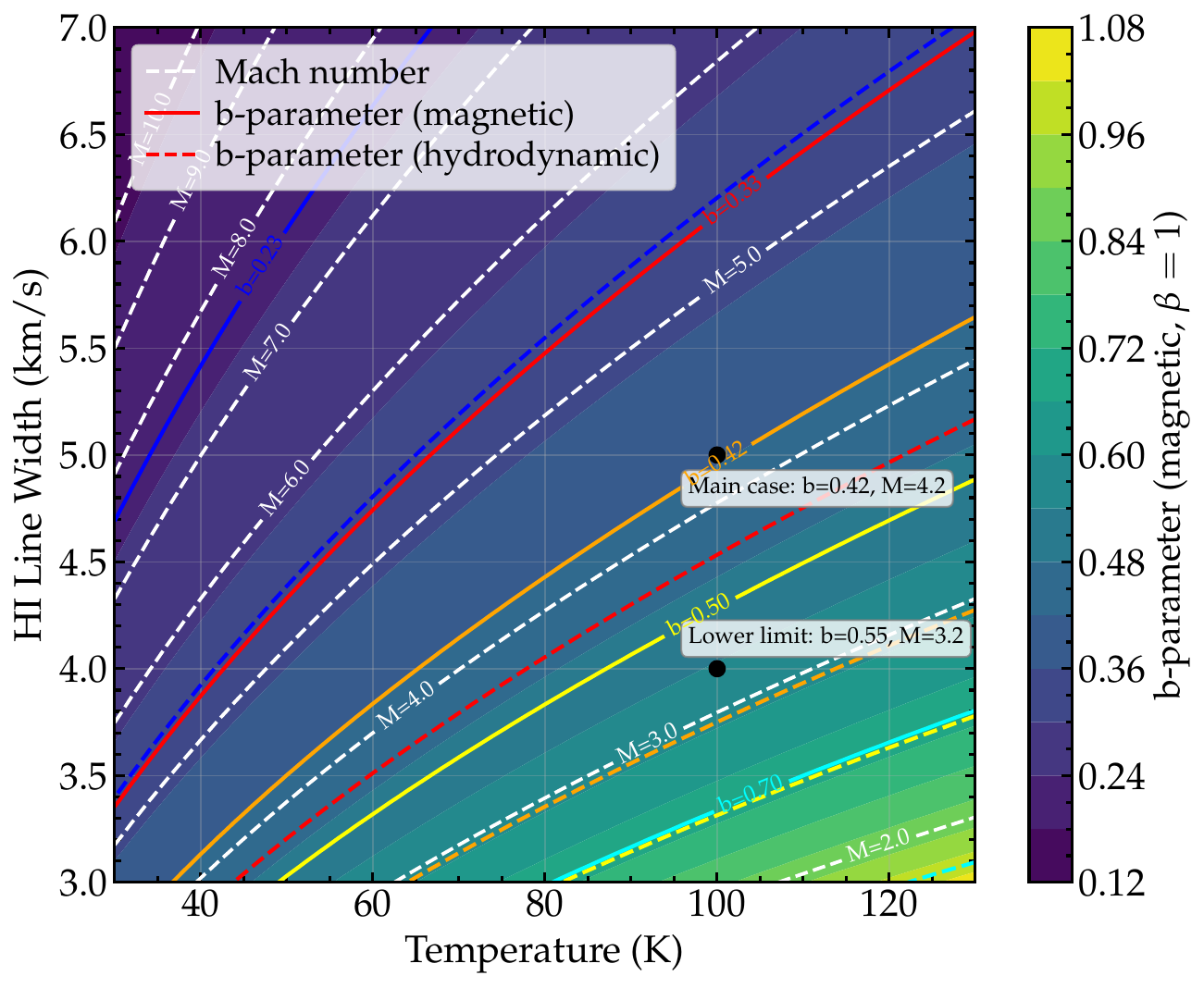}
\caption{Parameter space for the Mach number and $b$ parameter. The diagram displays the variation in $\mathcal{M}$ and $b$ as a function of the \HI\ line width and temperature for the HD case (dotted line) and the magnetic case (straight line).}
\label{space}
\end{figure}

\noindent We determined the sonic Mach number $\mathcal{M_{\mathrm{HI}}}$ for the \HI\ CNM phase from 
\begin{equation}\label{eq:mach-HI}
  \mathcal{M_{\mathrm{HI}}} = \frac{\sqrt{3}}{c_s} \, \sqrt{\frac{\Delta {\rm v_{HI}}^2}{8 \ln{2}} - \frac{k_B\,T} {\mu m_{\rm H}}} 
,\end{equation}
with the isothermal sound speed $c_s$ given as 
\begin{equation}\label{eq:cs-HI}
  c_s \,  = \sqrt{\frac{k_B T}{\mu m_{\rm H}}}
,\end{equation}

\noindent with the hydrogen mass $m_{\rm H}$, the Boltzmann constant $k_B$, and
the molecular weight $\mu$=1.27. 

For Draco, we obtain an average
\HI\ linewidth $\Delta {\rm v_{HI}}\sim$5 km s$^{-1}$ from the high-angular-resolution (1$'$) interferometric DRAO \HI\ data
\citep{Blagrave2017,Schneider2024}.  Note, however, that this is a
rough approximation. First, there is variation in the line profile
with sometimes two components, and second, the line widths for single
profiles varies between 4 and 7 km s$^{-1}$.  The temperature is
determined in different ways.  If we use the PDR toolbox
\citep{Pound2022}, with the Wolfire \& Kaufman model of 2020 as well
as with the KOSMA-tau model from 2020, with a CNM density of 50
cm$^{-3}$ as an upper limit (Sect.~\ref{subsec:prop}) and a radiation
field of $\chi \sim 2$ (G$_\circ \sim$3.4), we obtain a surface
temperature of $\sim$100 K. If we use Fig. 2 of \citet{Clark2019},
which presents magnetohydrodynamic (MHD) molecular cloud formation
simulations, we also obtain a gas temperature of $\sim$100 K. Note
that such a temperature is on the high end for cold CNM, but still
characterize CNM conditions.  With T = 100 K and $\Delta {\rm
  v_{HI}}\sim$5 km s$^{-1}$, we calculate a Mach number of
$\mathcal{M_{\rm HI}}$ = 4.1.  The width ($\sigma_{N}$) of the dust's
low N-PDF is 0.32 (Fig.~\ref{pdf}), so the
$b$ parameter for the hydrodynamic (HD) case is 0.29. If we consider a
magnetic field and assume a $\beta$ of 1, i.e. equipartition between
magnetic and thermal pressures, $b$ increases by a factor of
$\sqrt{2}$ to 0.42.  Such a value of the $b$ parameter is consistent
with fully developed turbulence. However, it is important to emphasize
that the value of b can exhibit significant variation depending on the
input parameters used in the calculations. For instance, an increase
in the temperature of the CNM or a decrease in the \HI\ line width
results in a lower Mach number but a correspondingly higher b-value.
Figure~\ref{space} illustrates how $\mathcal{M}$ and $b$ vary,
depending on the \HI\ linewidth and for a range of typical CNM
temperatures. From a physical perspective, the $b$ parameter is
expected to lie within the range from 0.33 to 1, with a typical value
around 0.5.  In summary, all reasonable variations within the
parameter space defined by line width and temperature lead to
supersonic Mach numbers, typically larger than 3, and yield $b$-values
close to 0.5. In all considered scenarios, the flow velocity exceeds
the thermal line width, aligning more closely with the relative
velocity between the IVC and the LVC.

We calculated the sonic Mach number $\mathcal{M}$ for the CO-bright H$_2$ gas phase from
\begin{equation}\label{eq:mach-H2}
  \mathcal{M} = \frac{\sqrt{3}}{c_s} \, \sqrt{\frac{\Delta {\rm v_{CO}}^2}{8 \ln{2}} - \frac{k_B\,T}{\mu m_{\rm CO}}} 
,\end{equation}
with the CO mass $m_{\rm CO}$ = 28~$m_{\rm H}$ and the mean molecular
weight in H$_2$ gas $\mu=2.33$.  We used a temperature of T = 12.5 K
for the molecular gas in Draco, which is an average from the
CO-determined values shown in Table~8 of \citet{Schneider2024} for a
density of a few times 1000 cm$^{-3}$, consistent with
\citet{Deschenes2017}.  The typical CO linewidth, $\Delta {\rm v_{\rm
    CO}}$, in Draco is 1-1.5 km s$^{-1}$ for $^{13}$CO and 2 km
s$^{-1}$ for $^{12}$CO for both the J=2$\to$1 and 1$\to$0 transitions
\citep{Mebold1985,Herbstmeier1993,Schneider2024,Bieging2024}.  We
derive a Mach number of $\mathcal{M}$ = 6.9 for $\Delta {\rm v_{\rm
    CO}}$ = 2 km s$^{-1}$ and $\mathcal{M}$ = 5.2 for $\Delta {\rm
  v_{\rm CO}}$ = 1.5 km s$^{-1}$.  Note that these values are valid
for the CO-bright H$_2$ phase and that $\mathcal{M}$ might be somewhat
smaller (and $b$ somewhat larger) in the more diffuse CO-dark
H$_2$. For the determination of the $b$ parameter, we again  used Eqs. 4
and 5 and a width of $\sigma_{N}$ = 0.34 from the \textit{Herschel}
N-PDF for the H$_2$ gas phase. We obtain $b$ = 0.20 (0.27) for $\Delta
{\rm v_{\rm CO}}$ = 2 (1.5) km s$^{-1}$ the HD  case and $b$
= 0.28 (0.37) for the magnetized case with $\beta$ = 1, corresponding
to mostly solenoidal driving. If the temperature is higher, the
Mach-number is lower and the $b$ parameter increases. However, we note
that the CO-linewidth is clearly not lower than 1 km s$^{-1}$ and
shows less variation than the \HI\ line width. 

Summarizing, the
cold \HI\ and the H$_2$ gas in Draco are both supersonic, with
$\mathcal{M}_{\rm HI} \sim$4 and $\mathcal{M}_{\rm H_2}\!\sim$5. These
values are consistent with what was obtained by \citet{Deschenes2017}.

We could then check to what extent the assumption of steady-state chemistry is fulfilled in Draco by calculating the chemical and dynamical timescales, $t_{chem}$ and $t_{dyn}$, respectively, following \citet{Wolfire2010}. In their turbulent scenario, the mass-weighted median density, $n_{med}$ is defined as 
\begin{equation}
n_{med}  = n \,\, \exp(0.5 \, \ln(1 + 0.25 \,\mathcal{M}^2)) \label{eq3} 
,\end{equation}

\noindent with the Mach number $\mathcal{M}$ and the density $n$. The chemical timescale is then \citep{Wolfire2010} 

\begin{equation}
t_{chem} = \frac{0.5}{n_{med} \,R}~, \label{eq4} 
\end{equation}

\noindent which is the time for atomic gas to become completely molecular and $R$=3 10$^{-17}$ cm$^{3}$ s$^{-1}$  
is the rate coefficient for H$_2$ formation on dust grains assuming solar metallicity \citep{Jura1974,Gry2002}. For the dynamical time for mixing material over the scale corresponding to an optical depth ${\rm A}_{\rm V}$, we used \citep{Wolfire2010}
\begin{equation}
t_{dyn} \, [\mathrm{s}] = \frac{L_{{\rm A}_{\rm V}}}{\sigma} = \frac{1.87 \, 10^{21} \, \cdot {\rm A}_{\rm V}}{10^5 \, n \, \sigma\, [\mathrm{km\,s^{-1}}]}   \label{eq5} 
\end{equation}
\noindent with the 1D velocity dispersion 
\begin{equation}
\sigma \, [\mathrm{km\,s}^{-1}] = \frac{\Delta {\rm v}\,[\mathrm{km\,s^{-1}}]}{\sqrt{(8 \, \ln(2))}}~.   \label{eq6} 
\end{equation}

This dynamical timescale describes the time needed for the turbulence
to mix gas over a length scale $L_{{\rm A}_{\rm V}}$ corresponding to
an optical depth ${\rm A}_{\rm V}$ to move molecular gas from the
dissociation front in a clump to the surface, and to bring atomic gas
from the surface to the interior. This scale is then divided by the
turbulent velocity field, characterized by $\sigma$ to obtain
$t_{dyn}$. With $\mathcal{M}$ = 4, $n$ = 50 cm$^{-3}$, \av = 0.33 and
$\Delta$v = 5 km s$^{-1}$ (Sect.~\ref{subsec:prop}), we obtain
$t_{dyn}$ = 1.87 Myr. Note that \citet{Bieging2024} calculated
dynamical timescales of individual features in Draco from their CO
observations. They obtained that a typical extended clump has a
transverse motion of $\sim$15$'$ in 10$^5$ yr.  For the chemical
timescale, following Eq.~\ref{eq4}, we estimate $t_{chem}$ = 3.3 Myr.
Even considering the uncertainties in all calculations, $t_{chem} >
t_{dyn}$ so that dynamical mixing effects cannot be neglected.  This
estimate shows that it is essential to consider dynamical effects and
apply time-dependent chemical models (gas + grains) on observational
data.

\subsection{Comparison of the N-PDF to turbulent molecular cloud formation simulations} \label{subsec:simu}

Various authors performing (M)HD turbulence
simulations
\citep{Glover2007,Glover2010,Micic2012,Gazol2013,Valdivia2016,Bialy2017}
argue that large density compressions caused by supersonically
turbulent gas allow H$_2$ to form more rapidly than in models without
turbulence.  It was shown that H$_2$ can be rapidly produced on
timescales of a few million years and that much of the H$_2$ is formed on very
small scales ($<$0.1 pc) in high-density gas, and then transported to
lower densities by turbulent motions. However, there are some subtle
differences in the predictions of these studies, which we outline
below and compare to our observational results.

\citet{Glover2007} perform turbulent box (size 20 pc) simulations and
investigate the H$_2$ formation timescale as a function of density. In
their models, the densities must be high in order to obtain a
significant H$_2$ fraction. For a density of 50 cm$^{-3}$ for Draco,
we find that 11\% of the gas is in molecular form (see also
\citealt{Schneider2022}), consistent with the $n=30$ cm$^{-3}$ case in
\citet{Glover2007} with only local shielding (that underestimates the
H$_2$ fraction).

\citet{Gazol2013}, \citet{Hennebelle2007}, \citet{Seifried2011}, and
\citet{Valdivia2016} study thermally bistable turbulent
flows. \citet{Audit2010} in addition compare 2-phase, isothermal and
polytropic flows.  All authors find double log-normal density and
column density PDFs for bistable flows, which is not surprising in
view of the bimodal nature of the distribution. Note, however, that
the two phases in these studies mostly correspond to WNM and CNM and
are thus not directly comparable to what we propose to see in Draco,
i.e. the cold atomic and molecular CNM phases. However,
\citet{Audit2010} noticed that their CNM PDF is not well described by
a purely log-normal distribution, suggesting that there can be
substructure in this part of the PDF; this could correspond to what we
observe in Draco.  \citet{Gazol2013} show N-PDFs as a function of Mach
number obtained from their purely HD thermally bistable
turbulent flows. Again, their N-PDFs are not purely log-normal and
their absolute peak values around a few 10$^{20}$ cm$^{-2}$ agree with
what we observe in Draco. However, it remains unclear which gas phases
(WNM or CNM) correspond to which part of their N-PDF.

In order to separate these phases, we calculated the N-PDFs from
existing simulations.  The first one is a MHD
simulation of the multi-phase turbulent ISM
\citep{Kritsuk2004,Kritsuk2017} and the second one a 3D zoom-in
simulation of the formation of two molecular clouds out of the
galactic ISM from the SILCC\footnote{SImulating the
LifeCycle of molecular Clouds;
http://www.astro.uni-koeln.de/$\sim$silcc.}-Zoom project
\citep{Walch2015,Girichidis2016,Seifried2017}.  Both simulations
include the earliest phases of cloud formation and are thus relevant
for the interpretation of the observed N-PDF of Draco. The advantage
of these simulations is that it is possible to perform PDFs of total,
atomic, and molecular hydrogen individually in different gas phases
(CNM, UNM, etc.). We do not expect a one-to-one correspondence since
the simulations setup was not designed for the Draco case, but
anticipate a qualitative resemblance.

\begin{figure}
\centering
\includegraphics[width=9cm, angle=0]{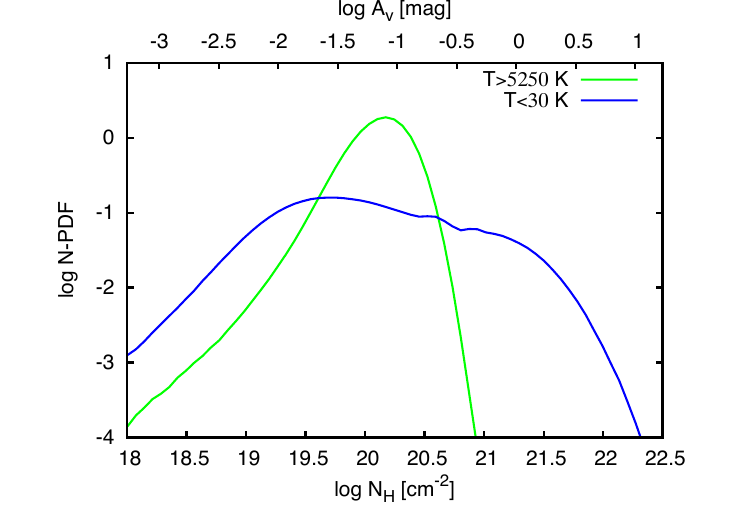}
\caption{PDF from a MHD simulation (magnetic field strength B=3$\mu$G) of forced multi-phase turbulence \citep{Kritsuk2004,Kritsuk2017} in a 200 pc box with a mean \HI\ density of 5 cm$^{-3}$ and a velocity dispersion of 16 km s$^{-1}$. The WNM is represented by the green line, the CNM (atomic and molecular) by the blue line.}
\label{alexei-pdf}
\end{figure}

\begin{figure}
\centering
\includegraphics[width=9cm, angle=0]{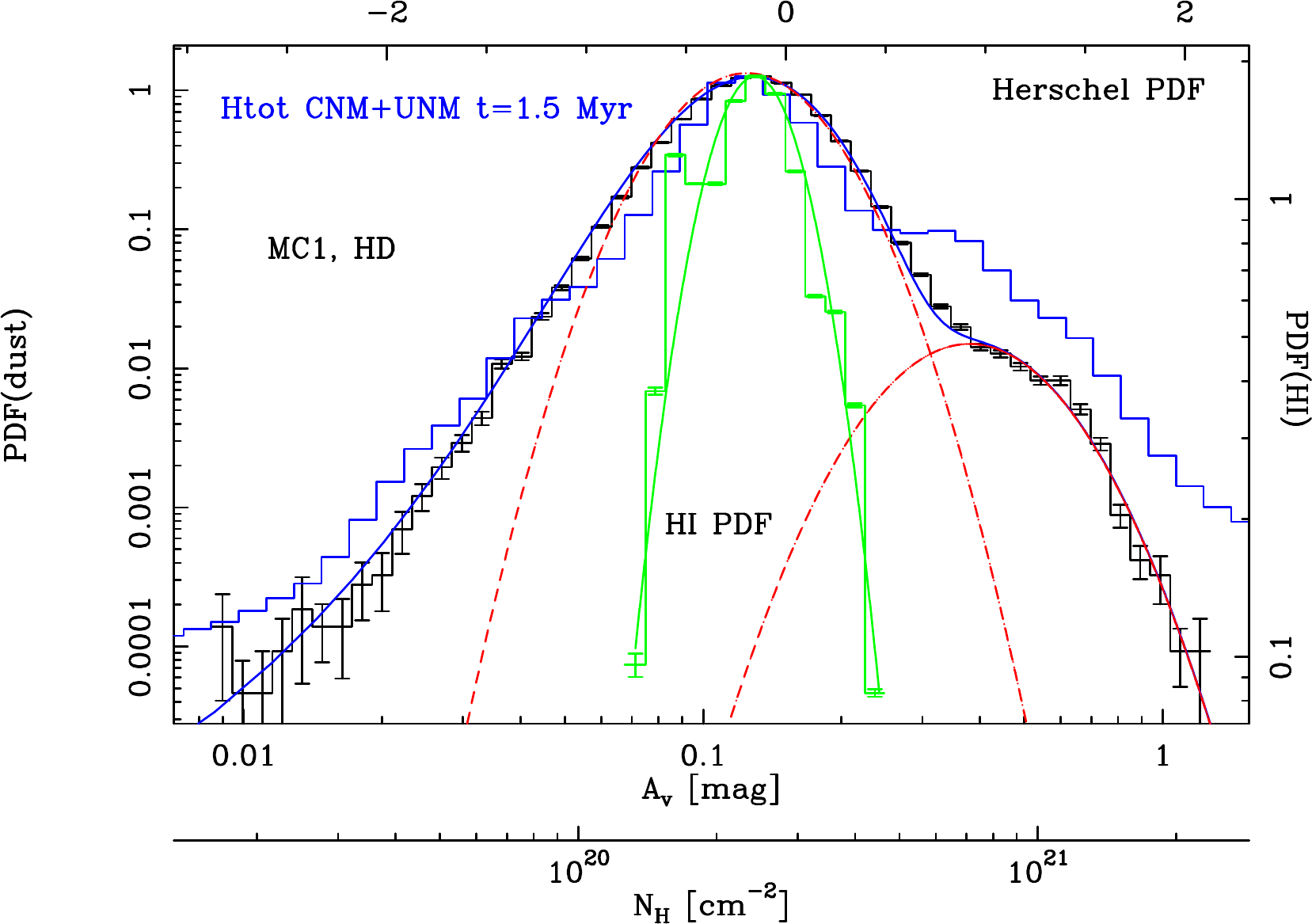}
\caption{PDFs from a SILCC-Zoom simulation together with the observed Draco PDF. The blue line shows the total hydrogen N-PDF at the 1.5 Myr timestep of the HD MC1 simulation \citep{Seifried2017}, which includes the CNM and UNM; see Sect.~\ref{sec:intro}. The black line is the observed N-PDF of Draco, the green line the \HI-PDF of Draco obtained from EBHIS, and the dashed red curves are the log-normal fits of the observed N-PDF. See also Fig.~\ref{pdf}.}
\label{daniel-pdf}
\end{figure} 

\subsubsection{Multi-phase MHD model} \label{model-1}
The simulation by Kritsuk et al. is a 200 pc multi-phase model with
uniform magnetic field strength of B = 3 $\mu$G, a mean \HI\ density
of 5 cm$^{-3}$, and a velocity dispersion of 16 km s$^{-1}$.  There is
no chemistry and no cooling below 18 K in this model, so model-derived
temperatures below 18 K may be wrong by a factor of $\sim$2.  The density
regime is fixed by the mean density of the model, so this should not
be expected to make an exact match to the observations. However, the
key physics (magnetized multi-phase turbulence) is included and allows
for a qualitative comparison to the Draco PDF. The model PDF is shown
in Fig.~\ref{alexei-pdf}. The blue line displays the CNM, the atomic,
and molecular phase. This is approximately what can be expected for
dust N-PDFs that are not sensitive to the high-temperature WNM.  The
green line displays the WNM PDF alone with the typical prominent peak
at low column densities. This particular simulation already reproduces
qualitatively a similar N-PDF (the blue line) as we observed for Draco
even though not all observational features match well (PDF width,
peak, etc.)

\subsubsection{SILCC model} \label{model-2}
The second model is based on the SILCC-Zoom project that simulates a
section of a galactic disc with solar neighbourhood properties and
solar metallicity at a size of 500 pc $\times$ 500 pc $\times \,\pm$5
kpc. The MHD and purely HD
simulations include (supernova-driven) turbulence and self-gravity; they
model the chemistry as well as the heating and cooling processes of the
ISM using the networks and methods described in \citet{Nelson1997},
\citet{Glover2007}, \citet{Glover2010}, and \citet{Glover2012}.
Various cloud structures develop from the diffuse ISM and are selected
for a dedicated high spatial and temporal resolution to study the
evolution of their internal substructure and chemistry. From various
zoom-in simulations, we selected the HD run using the
cloud `MC1' (extent 88$\times$78$\times$71 pc$^3$ and final mass 7.3
10$^4$ M$_\odot$) and focus on the early snapshots in the cloud
evolution, i.e. 1.5 and 2 Myr after the start of the zoom-in
procedure. At this time, the H$_2$ fraction of the MC1 cloud is
already significant ($\sim$50\% \citealp{Seifried2017}), and thus
probably more evolved than the Draco cloud\footnote{There is some
uncertainty concerning the molecular fraction in
Draco. \citet{Schneider2022} determined as a global average that
$\sim$10\% of the total gas in Draco is molecular, \citet{Moritz1998}
obtained spatially varying values between 0 and 50\%, and
\citet{Bieging2024} show the detailed distribution of molecular
fraction over the dense filaments, with 90\% only in the highest
column density areas.}, but still allows for a qualitative comparison
to our observations. 

Figure~\ref{daniel-pdf} shows the resulting
PDFs of the total hydrogen column density arising from the CNM and
UNM, as we expect for Draco, together with the Draco PDF for
comparison. Overall, we observe a good match between the observed and
simulated PDFs. The N-PDFs from the simulation also show a double-peak
at similar column densities as the Draco N-PDF. As said above, the
H$_2$ fraction is slightly higher than in the observed N-PDF, which is
not surprising because the MC1 cloud at these two time steps is most
likely more evolved as Draco.  Only the N-PDF from the simulations
including CNM and UNM reproduce our observed Draco N-PDF. The PDFs
produced from the same simulation setup but focussing only on CNM as
well as the PDFs from the H$_2$ phase only (CNM + UNM or CNM only) do
not match the Draco PDF. Interestingly, most of the MHD simulations
result in PDFs with a larger molecular fraction and do not reproduce
the observed Draco N-PDF.

\section{Conclusions} \label{sec:discuss}

Summarizing our results, we have a
scenario in which high-density (up to 10$^3$ cm$^{-3}$), cold
($\mathrm{T}\sim10-20$ K), and small molecular gas clumps with a low
filling factor \citep{Deschenes2017} are embedded in a more diffuse
atomic phase ($n<50$~cm$^{-3}$, $\mathrm{T}\sim 100$ K).  Cooling via
dust and atomic FIR lines, mainly the \CII\ 158 $\mu$m line and the
\CI\ 609 $\mu$m line \citep{Glover2015}, would also explain the nearly bimodal gas distribution, i.e. the atomic and
molecular CNM phase that also causes the double log-normal N-PDF.  The
expected \CII\ and \CI\ intensities, however, are estimated to be
extremely low \citep{Clark2019}.  On the other hand, if shocks are involved, the clump surfaces would be significantly heated and could then
be traced in \CII,\ for example. This is indeed the case, as shown
by \citet{Schneider2024}, who report the detection of the \CII\ line in
Draco at the position of dense clumps.

The dynamical scenario for H$_2$ formation
\citep[e.g.][]{Ball1999,Hartmann2001,Vaz2006,Heitsch2006,Banerjee2009,Audit2010,Clark2012,Valdivia2016,Seifried2017,Colman2025}
depicts a picture in which warm, trans-sonically turbulent \HI\ flows
are converging and the compression of \HI\ gas causes the gas to cool
via thermal instability to form high-density H$_2$
\citep[e.g.][]{Walder1998,Audit2005,Klessen2010}.  In Draco, the
scenario may be somewhat different because the Draco IVC has a
significant velocity of around $-23$~km~s$^{-1}$ and falls through the
WNM onto the Galactic disc. Regardless of whether the IVC has its origin
in extragalactic gas infall or from a Galactic fountain process, WNM
gas is compressed and cooling then leads to the nearly instantaneous
formation of atomic CNM and H$_2$ clumps.  In the cold, supersonic
H$_2$ phase, a 2-to-1 ratio between the energy in solenoidal and
compressive modes is established, so the estimated $b$ value for this
mixed case is $\sim$0.3-0.5 \citep{Federrath2010}, which is consistent with our
value of $b$.

Draco can thus not be considered a typical example
of colliding \HI\ flows, in particular because the existing `flows',
the HVC and IVC, have no velocity link and are well separated
(Fig.~\ref{draco-HI}). Moreover, Draco is probably a special type of
cloud in which the density structure and chemistry are influenced by
shocks that naturally occur during the passage of the IVC through the
WNM.  A similar scenario was suggested by \citet{Bieging2024} based
on simulations of \citet{Saury2014}, in which supersonic turbulence is
created as the infalling IVC encounters the WNM layer in the Galactic
plane, leading to a rapid formation of CO.

We propose that the shock compression leads to a large distribution of
small H$_2$ clumps embedded in the atomic phase. We note that the FIR
emission in high-density regions in Draco is unusually high, compared
to other IVCs, which could be an indication of shock heating.  This
was first seen in IRAS (Infrared Astroomical Satellite) 12 and 25 $\mu$m maps \citep{Odenwald1987} and
then in \textit{Herschel} flux maps
\citep{Deschenes2017,Schneider2024}. Again, \citet{Schneider2024}
demonstrate that shock excitation can be responsible for the high FIR
intensities.
  
We presume that Draco is seen in an early evolutionary phase and that
it will continue to grow in mass via accretion from the surrounding
atomic gas that becomes molecular at the stagnation point above a
certain extinction value, as suggested in the numerical simulations by
\citet{Heitsch2006} and \citet{Clark2012}.

\begin{acknowledgements}
N.S. and E.K. acknowledge support from the FEEDBACK plus project that was supported by the BMWI via DLR, Projekt Number 50 OR 2217. \\  
This work was supported by the Collaborative Research Center 1601 (SFB 1601 sub-projects A6, B1, B2, B4) funded by the Deutsche Forschungsgemeinschaft (DFG, German Research Foundation) – 500700252. \\
S.D. acknowledges support from the International Max Planck Research School (IMPRS) for Astronomy and Astrophysics  at the Universities of Bonn and Cologne. \\
A.K. was supported in part by the NASA Grant 80NSSC22K0724.
\end{acknowledgements}

\bibliography{draco.bib}

\begin{appendix} 

\section{The \textit{Herschel} column density map} \label{app-cib}

\begin{figure}
\centering
\includegraphics[width=9cm, angle=0]{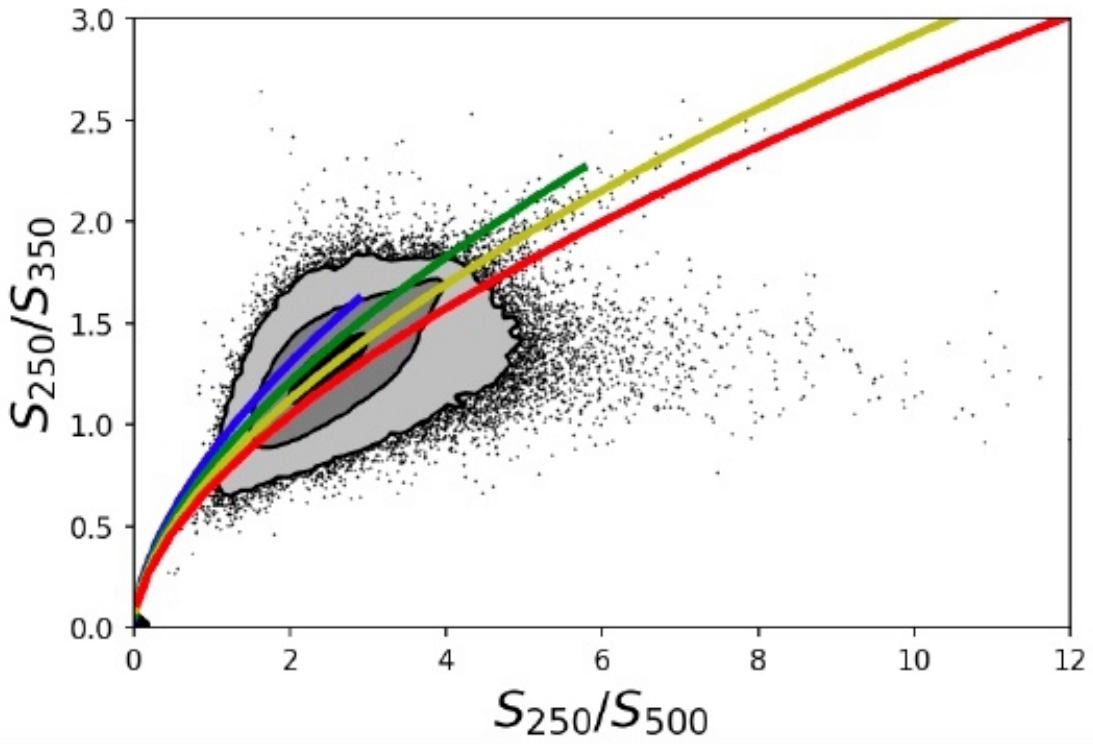}
\caption{Colour-colour plot of the \textit{Herschel} 250/350 $\mu$m vs 250/500 $\mu$m ratios. Blue, green, yellow, and red curves are theoretical colour-colour curves for a dust emissivity index $\beta$ = 0, 1, 2, and 3, respectively. Contours mark the 68\%, 95\%, and 99\% confidence levels.
}
\label{CIB}
\end{figure}

The \textit{Herschel} hydrogen column density map was derived following
the methodology described in \citet{Schneider2022}. This approach
involves a pixel-by-pixel grey-body fit to the spectral energy
distribution (SED), utilizing the \textit{Herschel} flux maps at 160,
250, 350, and 500 $\mu$m, all convolved to a common angular resolution
of 36$''$. This method is widely adopted within the astrophysical
community and has been extensively referenced
\citep{Andre2010,Russeil2013,Lombardi2014,Schneider2015a,Stutz2015,Pokhrel2020,Spilker2021,Schneider2022}.
All maps were subjected to absolute flux calibration. The {\sc
  zeroPointCorrection} task in HIPE was employed for SPIRE and IRAS
maps and PACS data. SPIRE maps were calibrated for extended
emission. The SED fitting procedure assumed a fixed specific dust
opacity per unit mass (dust+gas) characterized by the power law
$\kappa_\nu = 0.1\,(\nu/1000\,{\rm GHz})^{\beta}$~cm$^2$/g, with
$\beta = 2$, while leaving the dust temperature and column density as
free parameters \citep[see][for
  details]{Hill2011,Russeil2013,Roy2013}.  For Draco, calibration
uncertainties of 10\% for SPIRE and 20\% for PACS were applied to
weight data points at each wavelength. In \citet{Schneider2022}, an
alternative weighting using full noise maps was tested but was found
not to have an impact due to the high signal-to-noise ratio of
the \textit{Herschel} data. Draco is expected to exhibit diffuse cloud
conditions with minimal ice accretion or dust coagulation. Based on
\textit{Planck} observations, \citet{Juvela2015} derived a dust opacity
value of $\kappa_\nu = 2.16 \times 10^{-25}~{\rm cm}^{-2}/{\rm H}$ for
such regions, consistent with standard interstellar reddening, which
we adopted for Draco.

To assess potential contributions from the cosmic infrared background and unresolved extragalactic point sources, we constructed a
colour-colour diagram using the S250/S350 and S250/S500 ratios
(Fig.~\ref{CIB}).  The pixel distribution closely resembles typical
Galactic values, suggesting minimal contamination from extragalactic
sources or cosmic infrared background anisotropies. Potential contributions appear as
outliers, located within S250/S350 ratios of 1–1.5 and S250/S500
ratios of 6–12, with an estimated contribution of less than a few
percent. These results confirm that a dust emissivity index of $\beta
= 2$ provides an excellent fit to the data.  Opacity, dust
temperature, and $\beta$ maps of Draco, based on \textit{Planck} data,
were previously presented in \citet[see their Fig. 15]{Irfan2019},
showing significant variation depending on the method employed (e.g.
premise, GNILC, or 2013). \citet{Bieging2024} cite a study of
\citet{Singh2022} who used a SED fitting method with variable $\beta$
and obtained a total hydrogen column density map that is a factor of 3
lower than what is obtained by \citet{Schneider2022} and
\citet{Bieging2024}.  Summarizing, while a detailed comparison of
methods for deriving column density maps from \textit{Planck} and \textit{Herschel} is an important topic, it lies beyond the scope of this
study.  Using the simple conversion of the 250 $\mu$m \textit{Herschel}
flux map into a total hydrogen column density with N(H) = 2.49$\times$
10$^{20}$ I(250) [MJy sr$^{-1}$] H atoms/cm$^2$ \citep{Deschenes2017}
results in a column density map that is typically 30\% to 50\% higher
than what we obtained with the SED fit.  One reason could be that
because the calibration of the dust opacity is performed at long
wavelengths with \textit{Planck} data, typically at 353 GHz, a change of the
wavelength exponent $\beta$ affects the opacity and thus column
density estimation at smaller wavelength, in this case at 250
$\mu$m. We used $\beta$=2, in contrast to the $\beta$ of around 1.6 used in
Miville-Deschenes et al. Hence, the dust model assumed here has a
higher opacity at 250 $\mu$m, thus translating into a somewhat lower
column density on our map. We note also that relying solely on one
wavelength is inherently less reliable compared to a full SED fit
across multiple wavelengths. A detailed discussion of the
uncertainties related to the different methods how to obtain a column
density map in Draco is found in \citet{Bieging2024}.

\section{N-PDF} \label{app-pdf}

\begin{figure}
\centering
\includegraphics[width=9cm, angle=0]{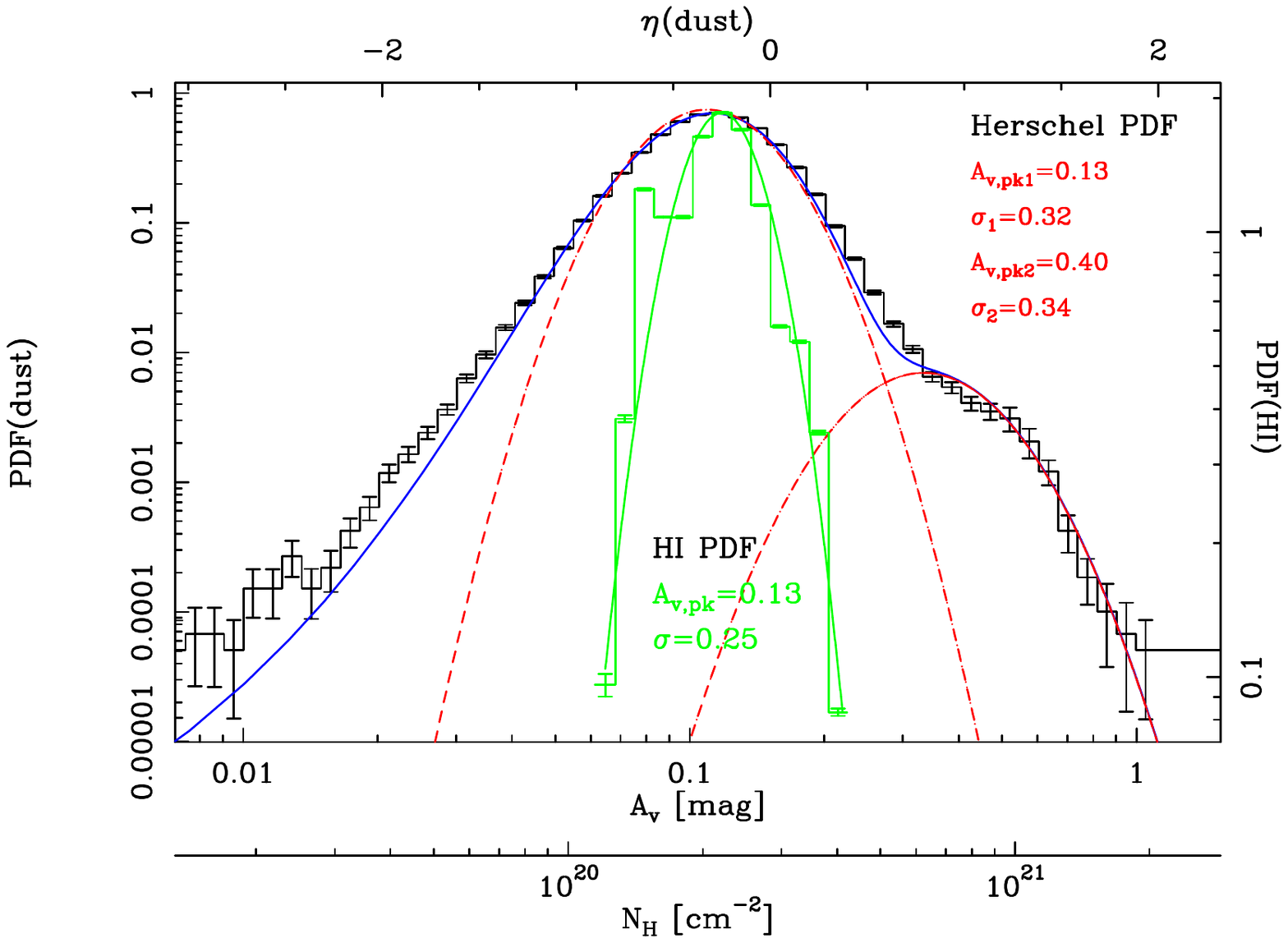}
\caption{N-PDF of Draco from \citet{Schneider2022}.  The black
  histogram indicates the N-PDF obtained from \textit{Herschel} data and
  the blue line its analytic description. The dashed red line gives
  the fit of two log-normal PDFs and takes into consideration the noise
  contribution that leads to a nearly linear behaviour at low column
  densities \citep{Ossenkopf2016}. The green histogram displays the
  N$_{\rm HI}$-PDF of the \HI\ data, and the continuous line a single
  log-normal fit. The fitted peak positions of the PDFs and the widths
  ($\sigma$ in units of $\eta$=ln(N/$\langle N \rangle$)) are given in
  the panel. Error bars are based on the method presented in
  \citet{Jaupart2022}. The left y-axis gives the probability value for
  the \textit{Herschel} map, and the right y-axis for the \HI\ map.}
\label{pdf}
\end{figure}

Figure~\ref{pdf} shows the N-PDF of Draco obtained from \textit{Herschel}
and \HI\ EBHIS data and presented in \citet{Schneider2024}. The shapes
of the PDFs are discussed in detail in \citet{Schneider2022}. We note
that the width of the \HI\ PDF will be systematically underestimated
because of the low angular resolution of 10$'$ of the \HI\ map
\citep{Schneider2015a,Ossenkopf2016}. The location of the peak of the
PDF, however, is unaffected by the spatial resolution (see the appendix in
\citet{Schneider2015a}.  We updated the PDF determination by
using the proper ergodic theory developed by \citet{Jaupart2022} for
the estimate of the uncertainties in a finite map. It compares the
size of correlated regions in the individual column density bins to
the total number of samples in that bin. We used two different
approaches to practically measure this size. In an isotropic
approximation, we fitted the correlation length, $l_\mathrm{corr}$, and
obtained the size as $\pi l_\mathrm{corr}^2$. In a discretized approach,
we counted the pixels in the 2D auto-correlation function
that fall above a level of $1/e$. In general, the results of the two approaches agree
well, within 10-20\,\%. To be conservative, we always used the larger of
the two values to compute the error bars here. It turns out that for
$A_\mathrm{V}$ values between 0.025 and 0.5, the error bars hardly
differ from those shown in \citet{Schneider2022}. The uncertainty
becomes somewhat larger for $A_\mathrm{V}> 0.7$, slightly modifying
our fit result but not changing any of the conclusions drawn in
\citet{Schneider2022}.

\end{appendix}

\end{document}